\documentclass{desyproc}

\begin{document}
\title{Polarization measurements and their perspectives: PVLAS Phase II}

\author{{\slshape G. Cantatore$^1$, R. Cimino$^2$, M. Karuza$^1$, V. Lozza$^1$, G. Raiteri$^1$}\\[1ex]
$^1$Universit\'a and INFN Trieste, Via Valerio 2, 34127 Trieste, Italy\\
$^2$INFN Lab. Naz. di Frascati, Via E. Fermi 40, 00044, Frascati, Italy}

\contribID{cantatore\_giovanni\_pvlasII}

\desyproc{DESY-PROC-2008-02}
\acronym{Patras 2008} 
\doi  

\maketitle

\begin{abstract}
We sketch the proposal for a ÒPVLAS-Phase IIÓ experiment. The main physics goal is to achieve the first direct observation of non-linear effects in electromagnetism predicted by QED and the measurement of the photon-photon scattering cross section at low energies (1-2 eV). Physical processes such as ALP and MCP production in a magnetic field could also be accessible if sensitive enough operation is reached. The short term experimental strategy is to compact as much as possible the dimensions of the apparatus in order to bring noise sources under control and to attain a sufficient sensitivity.
We will also briefly mention future pespectives, such as a scheme to implement the resonant regeneration principle for the detection of ALPs.
\end{abstract}

\section{Introduction}
The polarization of light plays an important role in many phenomena where photons interact with matter and other electromagnetic fields. It is well known from Quantum Electrodynamics (QED), and, more broadly, from particle physics, that photons impinging on a pneumatic vacuum region where an external magnetic field is present, are subject to various interactions at the microscopic level \cite{origQED}, such as photon-photon scattering \cite{photonphoton} and real, or virtual, production of Axion Like Particles (ALPs) and Mini Charged Particles (MCPs) coupled to two photons  \cite{origaxions}. By carrying out sensitive polarization measurements, it is possible to gain direct access to the physical quantities regulating these processes: photon-photon scattering and virtual particle production both result in an ellipticity acquired by the light beam, while real production results in a dichroism, that is an apparent rotation of the polarization plane.
Taking into account experimental parameters available in a laboratory, one must detect, as in the case of photon-photon scattering, acquired ellipticities of the order of $10^{-11}$ or less. It is apparent, then, the crucial need for a high sensitivity ellipsometer in all these measurements. This kind of experimental investigation was started at CERN by the group led by E. Zavattini \cite{propexpQED}. Later on, the BFRT experiment, based on an ellipsometer equipped with a multipass optical cavity and a pair of 4 T, horizontally placed, dipole magnets, reached a sensitivity just below $10^{-6} \ 1 / \sqrt{ \text {Hz}}$ \cite{BFRT}. BFRT set the first limits on the quantities involved in the physical processes described above, especially on particle production. The PVLAS experiment,  at the Legnaro National Laboratories of INFN, Italy, used a high-finesse  ($\approx$100000) Fabry-Perot (FP) optical resonator and a 5 T super-conducting dipole magnet, placed vertically and capable of rotating around its own axis, and has reached a sensitivity of $\approx 5 \cdot 10^{-7} \ 1 / \sqrt{ \text {Hz}}$ \cite{pvlas}.
The study of magneto-optical effects in vacuo is also presently under way by several groups of researchers, among which the BMV project at the University of Toulouse, France \cite{BMV}, the Q$\&$A experiment in Taiwan \cite{QandA}, and the OSQAR experiment at CERN \cite{pugnat}. All these efforts have common features: a low energy (1-2 eV), low flux ($\approx 10^{18}$ ph/s) photon beam probes a pneumatic vacuum region subject to a magnetic field, the sought-after effect is made time-varying in order to employ signal extraction techniques, and the length of the optical path in the magnetic region is amplified by means of an optical resonator. They also share a common problem: how to minimize the noise background.


\section{PVLAS Phase II}
After an effort lasting several years, the possibilities of the "phase I" of PVLAS can be considered exhausted. The positive signal published in 2006 has subsequently been proven to be an instrumental effect, and the best sensitivity reached is $\approx 5 \cdot 10^{-7} \ 1 / \sqrt{ \text {Hz}}$ \cite{pvlas}, meaning that detecting QED effects would take about 80 years of measurement time. In fact, the original PVLAS apparatus was not completely optimized from an optics point of view. The large dimensions of the granite tower holding the optical components were actually dictated by the necessity of enclosing a 4 m high, 5 ton cryostat, and not by an optimal optics design. During operation it has been found that the tower actually moves, transferring this movement to the optics, especially the resonator mirrors. For instance, the induced birefringence due to beam movement on mirror surfaces has been measured to be $\approx0.4 \ \text{m}^{-1}$. A sensitivity in ellipticity of  $5 \cdot 10^{-7} \ 1 / \sqrt{ \text {Hz}}$ then means that relative movement between the top and bottom optical benches must be  $< \ 2 \cdot 10^{-7} \ \text{m} /  \sqrt{ \text {Hz}}$.
This is impossible to achieve unless, perhaps, the entire PVLAS apparatus is rebuilt from scratch. With such a large apparatus it is also very hard to control overall thermal and acoustic noises. The main point we wish to make here, is that there is no reason of principle to keep a large optics tower.

The basic problem one must attack in order to progress with ellipsometric techniques is the reduction of the noise background. With this in mind we have started \textit{"PVLAS Phase II"}, based on the following ideas. Abandon the large optics tower and compact the apparatus down to table-top size, mounting all components on a single optical bench in a well tested vacuum system. Carefully characterize the apparatus step by step and implement from the start  all "passive" means of noise reduction, such as vibration isolation, environmental shields, solid and remotely controlled optics mounts. Reduce the number of optical components by developing a new ellipticity modulator which can be integrated on a resonator mirror. Use rotating permanent magnets in order to obtain a large duty cycle, fringe-field free operation, and the possibility to rotate the magnets at relatively large frequencies, up to 10-20 Hz.
We are at present operating, in the INFN Laboratories in Trieste, Italy, a prototype table-top ellipsometer with these characteristics: a frequency-doubled Nd:YAG CW laser emitting 900 mW at 1064 nm and 20 mW at 532 nm; a 50 cm long, 2.3 T permanent magnet; a 1 m long high-finesse FP resonator (a finesse of 170000 has been reached during tests). The vacuum system is capable of reaching a base pressure of $\approx 10^{-8} \ \text{mbar}$ and the composition of the residual gas can be dynamically sampled. Preliminary noise tests at atmospheric pressure, with the FP removed, indicate a promising sensitivity of $\approx 2 \cdot 10^{-8} \ 1 / \sqrt{ \text {Hz}}$ at 1 Hz, in slight excess of shot-noise limited operation. We envision a two-three year effort on a three step approach.
\\
\textit{Prototype ellipsometer (already existing)} - 900 mW - 1064 nm laser and 20 mW - 532 nm laser, stress-birefringence ellipticity modulator, 1 m long 220000 finesse FP, with 2.3 T -  50 cm long rotating permanent magnet, environmental screens, analog frequency-locking feedback on FP resonator.
\\
\textit{Advanced ellipsometer} - prototype plus intensity stabilization to reduce laser Residual Intensity Noise (RIN), stress-birefringence modulation directly on the cavity mirrors, lower noise electronic detection chain, improved acoustic isolation, digital frequency-locking feedback on FP resonator.
\\
\textit{Advanced ellipsometer with power upgrade} - power upgraded laser to 600 mW or more at 532 nm, light injection and extraction from the cavity via optical fiber.
\\
All three configurations could in principle be instrumented with a second 2.3 T, 50 cm long, permanent magnet to gain a factor 2 in signal and to allow zero-field equivalent measurements by crossing the two magnets at $90^{\circ}$. 

It is important to note that the table-top ellipsometer is not just a test apparatus to research noise sources, rather, if a good enough sensitivity is reached (at least $\approx 10^{-8} \ 1 / \sqrt{ \text {Hz}}$) it could actually achieve the first detection of the vacuum magnetic birefringence due to photon-photon scattering. Table \ref{tabletimes} gives the integration times, in units of 8-hour standard days, needed to achieve detection of QED ellipticity using a 220000 finesse FP resonator and one or two magnets. "IR" refers to 1064 nm, while "GREEN" refers to 532 nm. Also, the three ellipsometer advancement steps are considered. Notice the worst case of a measurement time of 188 days, which is difficult to achieve, but by no means impossible.

\begin{table}
\begin{tabular}{ccccccc}
\hline\hline
Config. & & IR & & GREEN\\\hline\hline
 & & Prot. & Adv. & Prot. & Adv. & Adv. pow. upg.\\ 
 & Sens. $ [1/ \sqrt{\text{Hz}}]$   & $10^{-8}$ &  $6 \cdot 10^{-10}$ & $10^{-8}$ & $6 \cdot 10^{-9}$ & $10^{-9}$\\ 
 One mag. & Meas. time \\
 & (8-hr. days)  & 188 &  0.675 & 47.1 & 16.9 & 0.471\\
 Two mag. & Meas. time \\
 & (8-hr. days)  & 47.1 &  0.169 & 11.7 & 4.2 & 0.12\\
\hline\hline
\end{tabular}
\caption{Minimum measurement times necessary to detect QED photon-photon scattering for several apparatus configurations (see text).}
\label{tabletimes}
\end{table}

\section{Future perspectives}

\begin{figure}[ht]
\centerline{\includegraphics[width=0.60\textwidth]{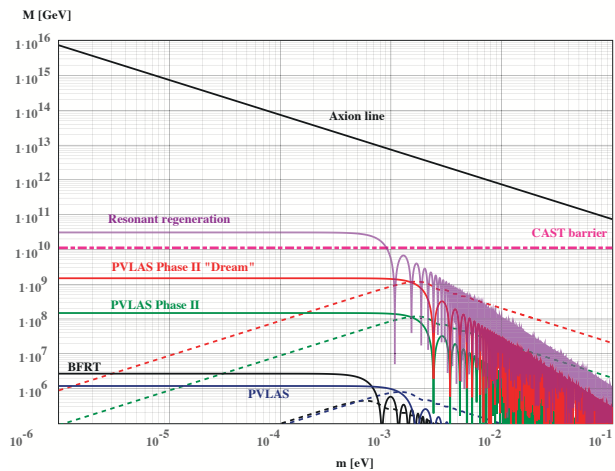}}
\caption{Upper bounds from laser experiments on ALP production. The CAST limit is shown, along with the axion line given by current models for QCD axions. An upper bound corresponding to successful measurement based on the resonant regeneration principle is also plotted (see text).}\label{Figure1}
\end{figure}

In addition to observing QED, a sensitive ellipsometer could be used, for instance, to set very stringent laboratory bounds on ALP mass and coupling constant to two photons. Figure \ref{Figure1} shows the mass [m]-inverse coupling [M] plane for ALPs. Upper bounds from completed/projected polarization experiment are shown along with the "axion line", roughly corresponding to the currently accepted models for QCD axions. Notice how, even in the best case, polarization measurements cannot reach the "CAST barrier", that is the limit obtained by the CAST helioscope at CERN \cite{cast}. 

A suggestive proposal has been put forth in Ref. \cite{resonant_reg}, where the authors envision a regeneration experiment with a FP cavity used to enhance ALP production in a first magnet, and a second FP enhancing ALP reconversion in a second magnet. The technical challenge is quite steep, involving the attainment of complete coherence of two high-finesse FP resonators.  However, using a scheme based on a double-wavelength emitting laser \cite{cantatore_talk}, we believe that such a goal is ultimately reachable. The graph in Figure \ref{Figure1} reports the bound which could be set by a successful resonant regeneration measurement. In our view this represents the only hope by a laboratory laser experiment of beating the "CAST barrier" .




\begin{footnotesize}



%

\end{footnotesize}


\end{document}